\documentclass[12pt]{article}

\usepackage{epsf,a4}

\title{Self-diffusion coefficients of charged particles: \\
       Prediction of
       Nonlinear volume fraction dependence 
       }
\author{M. Watzlawek\thanks{
              Present address: 
              Institut f\"ur Theoretische Physik II, 
              Heinrich--Heine--Universit\"at, Universit\"atsstr. 1, 
              D--40225 D\"usseldorf, Germany. 
              Electronic address: martin@thphy.uni-duesseldorf.de}
        and G. N\"agele \hfill \\ \hfill \\
        {\small Fakult\"at f\"ur Physik, Universit\"at Konstanz,} \\
        {\small Postfach 5560,
        D--78434 Konstanz, Germany}}

\date{December 6, 1996 \\
      {\small [published in Phys. Rev. E {\bf 56}, 1258 (1997)]}}

\begin{document}

\maketitle

\begin{abstract}
   We report on calculations of the translational and rotational
   short-time self-diffusion coefficients $D^t_s$ and $D^r_s$ for 
   suspensions of charge-stabilized colloidal spheres. These diffusion 
   coefficients are affected by electrostatic forces and many-body
   hydrodynamic 
   interactions (HI). Our computations account for both two-body and
   three-body HI. For strongly charged particles, we predict interesting
   nonlinear scaling relations $D^t_s\propto 1-a_t\phi^{4/3}$ and
   $D^r_s\propto 1-a_r\phi^2$ depending on volume fraction $\phi$, with
   essentially charge-independent parameters $a_t$ and $a_r$.
   These scaling relations are strikingly different from the corresponding
   results for hard spheres.
   Our numerical results can be explained using a model of
   effective hard spheres.
   Moreover, we perceptibly improve the known result for $D^t_s$ of hard 
   sphere suspensions.

   \vspace{1ex}

   \vspace{1ex}
   
\end{abstract}

Self-diffusion of spherical colloidal particles has been
studied experimentally over a wide range of time scales by means of
various scattering techniques, in particular by polarized and depolarized
dynamic light scattering (DLS). At short times on the scale
of DLS, the particles have only moved a small
fraction of their diameter $\sigma$, and the particle motion is determined
by solvent-mediated many-body hydrodynamic interactions (HI) weighted by the
equilibrium microstructure. The latter is determined by direct potential forces
arising, e.g., for hard-sphere particles from the steric repulsion between
the particles, and, in the case of charge-stabilized particles, from the 
electrostatic repulsion of overlapping double layers \cite{Pusey:91}.
The configuration-averaged effect of HI gives rise to values of the
translational and rotational diffusion coefficients $D^t_s$ and $D^r_s$
that are 
smaller than their respective Stokesian values at infinite dilution, i.e.
$D^t_0=k_BT/(6\pi\eta a)$ and $D^r_0=k_BT/(8\pi\eta a^3)$. Here,
$a$ is the particle radius, and $\eta$ is the shear viscosity
of the suspending fluid.

The properties of hard spheres are in various respects
easier to describe quantitatively than those of charge-stabilized
particles. As a consequence, there are many experimental 
\cite{Pusey:Megen:83,Degiorgio:95,Veluwen:87} 
and theoretical \cite{Degiorgio:95,Cichocki:88,Beenakker:Mazur:83}
results available on the short-time self-diffusion
coefficients of hard spheres. With regard to the computation of the first and
second virial coefficients of $D^t_s$ and $D^r_s$ in an expansion in terms
of the volume fraction $\phi$, the currently established results for the
normalized diffusion coefficients $H^t_s$ and $H^r_s$ are given by
\cite{Degiorgio:95,Beenakker:Mazur:83}
\begin{eqnarray} 
   \label{hts.hs}
   H^t_s&=&\frac{D^t_s}{D^t_0}=1-1.831\phi+0.88\phi^2+{\cal O}(\phi^3),
\\
   \label{hrs.hs}
   H^r_s&=&\frac{D^r_s}{D^r_0}=1-0.630\phi-0.67\phi^2+{\cal O}(\phi^3).
\end{eqnarray}
The possibility to expand $H^t_s$ and $H^r_s$ in
powers of $\phi$ arises from the fact that hard-sphere suspensions at small
$\phi$ are dilute both with respect to HI and to the microstructure.

While the short-time dynamics of hard spheres is well understood,
far less is known thus far about charge-stabilized suspensions. The
purpose of this letter is to show that there are striking differences in the
$\phi$-dependence of $H^t_s$ and $H^r_s$ between charged and uncharged
suspensions, and also to provide quantitative predictions.
These unexpected differences are most pronounced
for deionized, i.e. for salt-free suspensions of charged particles.
For such systems, our numerical results for $H^t_s$ and $H^r_s$ are well
represented by the parametric form $1+p\phi^\alpha$, where $\alpha$ is an
exponent larger than one. Due to the
strong direct interparticle interactions, 
deionized suspensions especially exhibit
pronounced spatial correlations even for very small $\phi$, say
$\phi\le10^{-4}$, so that
contrary to hard spheres these systems are dilute only with regard to HI.
The corresponding radial distribution function (rdf) $g(r)$ has a well
developed first maximum and it exhibits a so-called correlation hole, i.e.
$g(r)$ is essentially zero up to a well-defined nearest-neighbor separation
larger than $\sigma$ \cite{Naegele:Habil:published}.  
In comparison, the rdf of hard spheres is nearly equal to 
a unit step function $\Theta(r-\sigma)$ for $\phi\le0.05$, and an
analytical expression for $g(r)$ of hard spheres is known up to first order
in $\phi$ \cite{Kirkwood:35}.
Therefore, the calculation of $H^t_s$ and $H^r_s$ at small $\phi$ is more 
demanding for charged suspensions, because it is necessary to use static
distribution functions generated by integral equation methods or
computer simulations.

We base our calculations of $H^t_s$ for charge-stabilized suspensions on the
general expression
$
   H^t_s=\langle
            \mbox{Tr}\,{\bf D}^{tt}_{11}({\bf r}^N)
         \rangle/(3D^t_0)
$
as derived from the generalized Smoluchowski equation 
\cite{Jones:Pusey:91}. The corresponding expression for $H^r_s$ is obtained  
by replacing the superscipt
$t$ by $r$. The hydrodynamic diffusivity tensor 
${\bf D}^{tt}_{11}({\bf r}^N)$ (${\bf D}^{rr}_{11}({\bf r}^N)$)
relates the force (torque) exerted by the solvent on an arbitrary
particle $1$ with its translational (rotational) velocity.
$\mbox{Tr}\,{\bf D}^{tt}_{11}$ denotes the trace of 
${\bf D}^{tt}_{11}$, and the factor $1/3$ accounts for spatial isotropy.
Due to the many-body character of HI, both tensors depend on the
instantaneous $N$-particle configuration 
${\bf r}^N=({\bf r}_1,\ldots,{\bf r}_N)$,
and in principle the full $N$-particle distribution function is needed to
perform the equilibrium ensemble average $\langle\ldots\rangle$.
Thus, it is not possible to perform an exact calculation of
$H^t_s$ and $H^r_s$, that is valid for all particle concentrations. For small
$\phi$, however, when the mean particle distance gets sufficiently large,
a good approximation for these quantities is obtained 
by considering only two-body and, to leading order, three-body
contributions to the HI. For this reason, we use a rooted cluster 
expansion for the calculation of 
$H^t_s$ \cite{Degiorgio:95,Jones:1:88}, leading to the
following series expansion of $H^t_s$
\begin{equation} \label{hts.series}
   H^t_s=1+H^t_{s1}\phi+H^t_{s2}\phi^2+\ldots\ ,
\end{equation}
which we truncate after the third term.
Here, $H^t_{s1}$ is given in terms of integrals
\begin{equation} \label{hts1}
   H^t_{s1}=\frac{1}{a^3}\int_{2a}^\infty dr~r^2g(r)6\pi\eta a
               \left(
                  \alpha^{tt}_{11}(r)+2\beta^{tt}_{11}(r)
               \right),
\end{equation}
involving $g(r)$ and scalar two-body mobility functions
$\alpha^{tt}_{11}(r)$ and $\beta^{tt}_{11}(r)$, whose
expansions in powers of $(a/r)$ are known, in principle, 
up to arbitrary order
\cite{Jones:Schmitz:88,Cichocki:Felderhof:Schmitz:88}.
In our calculations, we include contributions to $\alpha^{tt}_{11}$ and
$\beta^{tt}_{11}$ up to ${\cal O}(r^{-20})$. 
The coefficient $H^t_{s2}$ is far more difficult
to calculate since it involves three-body HI. By considering the leading
term in the far-field expansion of the three-body part of 
${\bf D}^{tt}_{11}$ \cite{Beenakker:Mazur:83}, $H^t_{s2}$ is approximated by 
the threefold integral
\begin{eqnarray}
   \label{threebodytrans.one}
   H^t_{s2}&=&\frac{225}{64}\int_0^1~dt_{12}\int_0^1~dt_{13}\int_{-1}^1~d\xi~
            g^{(3)}(t_{12},t_{13},\xi) f_t(t_{12},t_{13},\xi),
\\ 
   \nonumber
   f_t(t_{12},t_{13},\xi)&=&\frac{t_{12}t_{13}}{h^{7/2}}~\xi
        \Bigg\{
           11t_{12}^2t_{13}^2-2\left(t_{12}^4+t_{12}^4\right)
           -10\xi t_{12}t_{13}\left(t_{13}^2+t_{12}^2\right)
\\
   \nonumber
        &+&\xi^2\left(10t_{12}^2t_{13}^2+6\left(t_{13}^4+t_{12}^4\right)\right)
           -6\xi^3t_{12}t_{13}\left(t_{13}^2+t_{12}^2\right)
           +3\xi^4t_{12}^2t_{13}^2   
        \Bigg\},
\end{eqnarray}
with $h(t_{12},t_{13},\xi)=t^2_{12}+t_{13}^2-2\xi t_{12}t_{13}$.
This integral involves the static triplet correlation function 
$g^{(3)}$ expressed in terms of $t_{12}=2a/r_{12}$, $t_{13}=2a/r_{13}$, and
$\xi={\bf r}_{12}\cdot{\bf r}_{13}/(r_{12}r_{13})$, where 
${\bf r}_{ij}={\bf r}_{i}-{\bf r}_{j}$ is the relative vector between the
particles $i$ and $j$, and $r_{ij}$ is its magnitude.

A similar analysis is used by us for calculating $H^r_s$, 
leading to expressions
for the coefficients $H^r_{s1}$ and $H^r_{s2}$, which appear in a
series similar
to Eq. (\ref{hts.series}), and which involve now rotational
two-body and three-body mobility functions. For conciseness,
we will not quote here the expressions for $H^r_{s1}$ and $H^r_{s2}$, since
these are given in Ref. \cite{Degiorgio:95}. 
Once again, we account for terms up to ${\cal O}(r^{-20})$ in the far-field 
expansion for the two-body mobility functions, and for the leading 
three-body part of ${\bf D}^{rr}_{11}$.

For charge-stabilized suspensions, it is only necessary to account for the
first few terms in the expansion of the two-body mobility functions,
since the integrals in Eq. (\ref{hts1}) converge rapidly because
$g(r)$ is practically zero at small $r$
\cite{Naegele:Habil:published}. On the other hand, many terms are needed for
hard spheres to accurately obtain the first virial coefficients
as depicted in Eqs. (\ref{hts.hs}) and (\ref{hrs.hs}).
Notice that the second virial coefficients for hard spheres are made up
of two contributions. The first one is due to $H^t_{s1}$ and $H^r_{s1}$,
with $g(r)$ expanded up to first order in $\phi$, whereas the second one 
arises from three-particle HI as embodied in $H^t_{s2}$ and $H^r_{s2}$
\cite{Degiorgio:95,Beenakker:Mazur:83}.
The second virial coefficient of $H^r_s$ in Eq. (\ref{hrs.hs})
was obtained by essentially accounting
for all two-body contributions in $H^r_{s1}$, and also for the
leading three-body contribution \cite{Degiorgio:95}. On the other hand,
only two-body terms up to ${\cal O}(r^{-7})$ plus the leading three-body term
were used so far in calculating the second virial coefficient of $H^t_s$, as 
given by the value $0.88$ in Eq. (\ref{hts.hs}) \cite{Beenakker:Mazur:83}. 
By considering terms up to
${\cal O}(r^{-20})$ in calculating $H^t_{s1}$, we obtain an improved
value of $-1.096$ for the two-body part of the second virial coefficient. 
Together with the three-body contribution $H^t_{s2}=1.81$, which is
obtained by Monte-Carlo integration of Eq. (\ref{threebodytrans.one}), we
get the improved result
\begin{equation} \label{hts.hs.better}
   H^t_s=1-1.831\phi+0.71\phi^2.
\end{equation}
This result
is in better agreement with experimental data
\cite{Degiorgio:95,Veluwen:87}
for hard-sphere suspensions than Eq. (\ref{hts.hs}).
The experimental data in Ref. \cite{Veluwen:87} especially
agree almost perfectly with Eq. (\ref{hts.hs.better}).

However, for charge-stabilized suspensions, it is not possible to use low-order
virial expressions of the static distribution functions.
We use instead results for $g(r)$, obtained from the rescaled
mean spherical approximation (RMSA), as applied to the one-component
macroion fluid model (OCM) of charge-stabilized
suspensions \cite{Naegele:Habil:published}. 
In the OCM, the effective pair potential $u(r)$ acting between two
particles consists of a hard-core part with diameter $\sigma$, and of a 
screened Coulomb potential 
$
   \beta u(r)=K\sigma\exp\left[
                            -\kappa(r-\sigma)
                         \right]/r
$
for $r>\sigma$. Here, $K=(L_B/\sigma)Z^2(1+\kappa\sigma/2)^{-2}$,
$L_B=e^2/(\epsilon k_BT)$, $\epsilon$ is the dielectric constant of the
solvent, and $Z$ is the effective charge of a particle in units of the
elementary charge $e$. $\kappa$ is given by the Debye-H{\"u}ckel relation
$\kappa^2=L_B\left[24|Z|\phi/\sigma^3+8\pi n_s\right]$, where $n_s$
is the number density of added 1--1-electrolyte, and the counterions
are assumed to be monovalent \cite{Naegele:Habil:published}.
For computing $H^t_{s2}$ and $H^r_{s2}$, $g^{(3)}(\bf{r},\bf{r}')$ is
needed as static input. To this end, we use for simplicity Kirkwood's
superposition approximation for $g^{(3)}(\bf{r},\bf{r}')$,
with the rdf calculated in RMSA. The threefold
integrals 
are calculated using a Monte Carlo method.

Since the observed qualitative differences in the 
short-time self-diffusion coefficients of charged and uncharged particles
are most pronounced for deionized charged suspensions, we concentrate
here on the case $n_s=0$. The system parameters used in our calculations
are typical for suspensions that have been under experimental
study \cite{Degiorgio:95,Bitzer:private}. 
If not stated differently, two-body contributions to
HI including terms up to ${\cal O}(r^{-20})$ are considered together
with the leading three-body contribution.
\begin{figure}[h]
   \epsfxsize=12.5cm
   \epsfysize=9.5cm
   \hfill\epsfbox{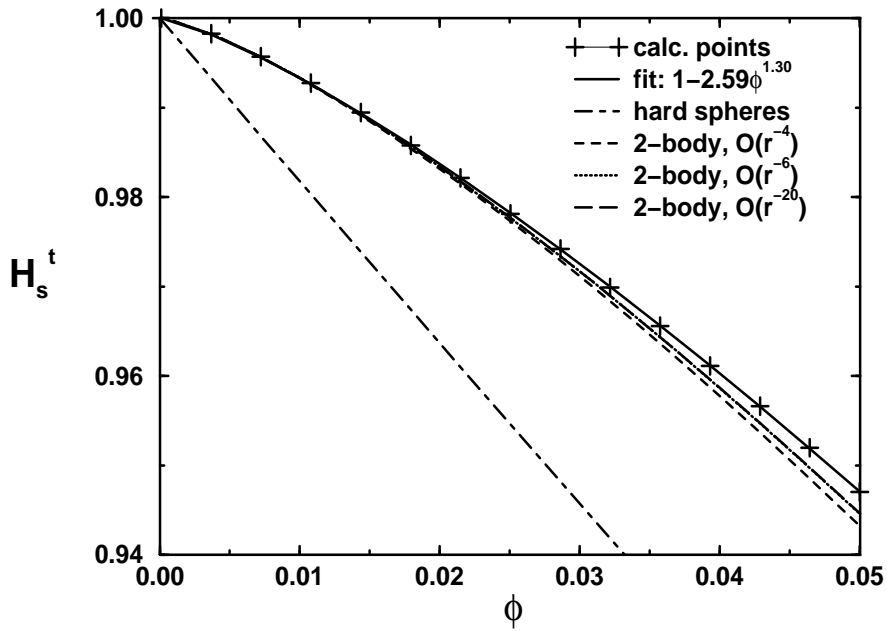}\hfill~
   \caption{ \label{hts.plot} \small
            $H^t_s$ versus $\phi$ for a deionized charge-stabilized
            suspension with $\sigma=90$nm, $Z=200$, $T=294$K, and 
            $\epsilon=87.0$. Solid line: best fit of the numerical results
            shows fractional $\phi$-dependence, i.e.
            $H^t_s=1-2.59\phi^{1.30}$, with exponent $\simeq 4/3$.
            Also shown is the dependence of $H^t_s$ on various two-body
            contributions to the HI. Included terms of the two-body
            expansion of ${\bf D}^{tt}_{11}$ as indicated in the figure.
            Dashed-dotted line: result for hard spheres
            according to Eq. (\ref{hts.hs.better}).
           }
\end{figure}
\begin{figure}[h]
   \epsfxsize=12.5cm
   \epsfysize=9.5cm
   \hfill\epsfbox{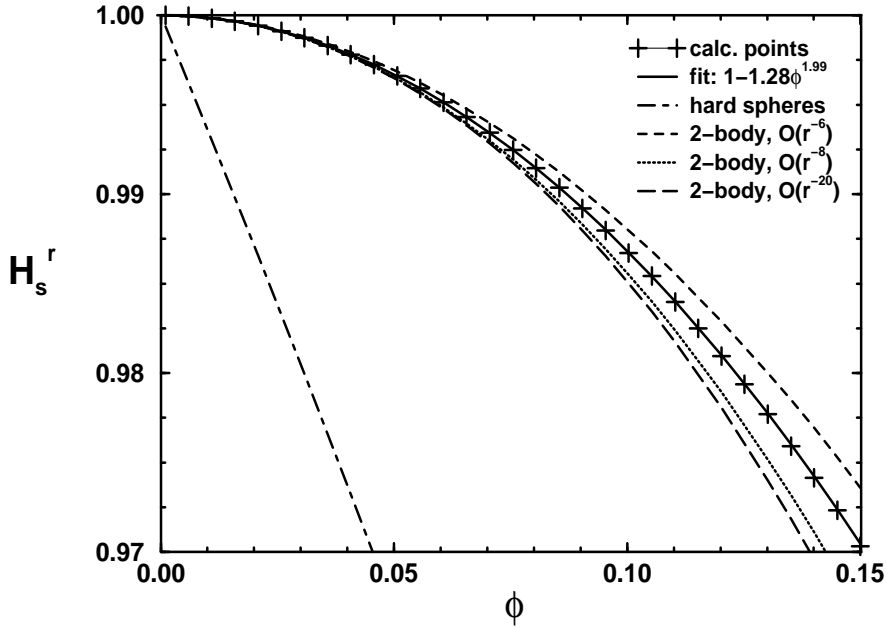}\hfill~
   \caption{ \label{hrs.plot} \small
            Results for $H^r_s$ obtained
            for a system with parameters as in Fig. \ref{hts.plot}, and
            compared with the corresponding result for hard spheres 
            given in Eq. (\ref{hrs.hs}).
            Best fit of the calculated points (solid line)
            has nearly quadratric $\phi$-dependence, i.e.
            $H^r_s=1-1.28\phi^{1.99}$, which extends to surprisingly
            large $\phi$.
            Further shown is the dependence of $H^r_s$ on various 
            terms of the two-body series
            expansion of ${\bf D}^{rr}_{11}$.
           }
\end{figure}
Figs. \ref{hts.plot} 
and \ref{hrs.plot} show our results for $H^t_s$ and $H^r_s$
as functions of $\phi$ (crosses). The corresponding results for hard spheres
are also included in these figures. Evidently, the effect of HI on 
$H^t_s$ and $H^r_s$ is less pronounced for charged suspensions.
Moreover, we find a qualitatively different $\phi$-dependence of $H^t_s$ and
$H^r_s$ for charged and uncharged particles. Whereas for hard spheres
the $\phi$-dependence of $H^t_s$ and
$H^r_s$ is linear at small $\phi$, we obtain for charged particles, from
a least-square fit of our numerical results to the form
$1+p\phi^{\alpha}$, the following interesting results
\begin{eqnarray}
   \label{hts.charged}
   H^t_s&=&1-a_t\phi^{1.30}, \ \ \ a_t=2.59
\\
   \label{hrs.charged}
   H^r_s&=&1-a_r\phi^{1.99}, \ \ \ a_r=1.28
\end{eqnarray}
with exponents close to $4/3$ and $2$, respectively.
Eq. (\ref{hts.charged}) is valid for $\phi\le0.05$, whereas from 
Fig. \ref{hrs.plot} it is seen that Eq. (\ref{hrs.charged}) is valid
even up to $\phi\le0.15$. The prefactors $a_t$ and $a_r$ are found to be 
nearly independent of $Z$ for $Z\ge200$. This
fact is illustrated in Fig. \ref{hts.charge.plot}, which
shows results for $H^t_s(\phi)$ for various values of $Z$.
\begin{figure}[h]
   \epsfxsize=12.5cm
   \epsfysize=9.5cm
   \hfill\epsfbox{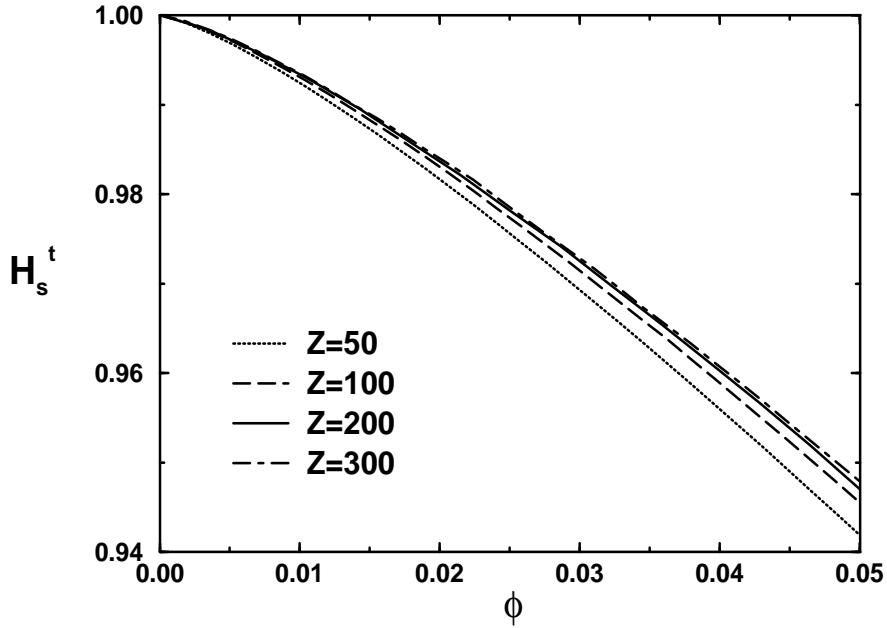}\hfill~
   \caption{ \label{hts.charge.plot} \small
            $H^t_s$ versus $\phi$ for various values of the effective 
            charge number $Z$ as indicated in the figure. All system
            parameters except $Z$ as in Fig. \ref{hts.plot}.
            Notice that $H^t_s$ becomes nearly independent of $Z$ for
            $Z\ge200$.
           }
\end{figure}
Notice that due to the $Z$-independence of $H^t_s$ and $H^r_s$, the
same Eqs. (\ref{hts.charged}) and (\ref{hrs.charged}) are recovered when
the accurate, but elaborate, Rogers--Young integral equation scheme
\cite{Naegele:Habil:published} is used for $g(r)$ instead of the RMSA.

We will now show that the 
occurance of exponents close to $4/3$ and $2$, and the $Z$-independence
of $a_t$ and $a_r$ can be understood in terms of a model
of effective hard spheres (EHS) with density-dependent effective diameter
$\sigma_{EHS}>\sigma$, which accounts for the extension of the correlation
hole. We can identify $\sigma_{EHS}=r_m$, where $r_m$ is the position of 
the principal peak of $g(r)$. It is now crucial to note for deionized 
suspensions
that $r_m$ as obtained from the RMSA coincides within $3\%$ with the
the average
geometrical distance $\bar{r}=\sigma(\pi/(6\phi))^{1/3}$ of two spheres. Thus,
we have the scaling relation
$r_m\propto\bar{r}\propto\phi^{-1/3}$. Here it is important that $Z$ be 
chosen large 
enough that the physical hard core of a particle is completely masked
by the electrostatic repulsion \cite{Naegele:Habil:published}. 
We now approximate $g(r)$ by the rdf $g_{EHS}(r;\phi_{EHS})$ of the
EHS model, evaluated at the effective volume fraction 
$\phi_{EHS}=\phi(\sigma_{EHS}/\sigma)^3$. When this approximation
for $g(r)$ is used, and if only the leading terms in the series expansions
of the two-body moblility functions are retained, we obtain the results
$H^t_s=1-a_t\phi^{4/3}$ and $H^r_s=1-a_r\phi^2$ with exponents very close to
our numerical results. Here
\begin{eqnarray} 
   a_t&=&\frac{15}{8}\phi^{-1/3}_{EHS}\int_1^\infty dx~
            \frac{g_{EHS}(z;\phi_{EHS})}{x^2}
       =\frac{15}{16}\phi^{-1/3}_{EHS}\int_0^\infty dz~
            z^2G_{EHS}(z;\phi_{EHS}),
\\
   \nonumber
   a_r&=&\frac{15}{16\phi_{EHS}}\int_1^\infty dx~
            \frac{g_{EHS}(z;\phi_{EHS})}{x^4}
       =\frac{15}{384\phi_{EHS}}\int_0^\infty dz~
            z^4G_{EHS}(z;\phi_{EHS}),
\end{eqnarray}
and $G_{EHS}(z)$ is the Laplace transform of $xg_{EHS}(x)$ with 
$x=r/\sigma_{EHS}$. Notice that $\phi_{EHS}$, and hence $a_t$ and $a_r$,
are independent of $\phi$ and $Z$ ($\ge200$) when $\sigma_{EHS}$ is 
identified as $r_m$. To obtain a rough estimate of $a_t$ and $a_r$, we can
further approximate $g_{EHS}(x)$ by $\Theta(x-1)$,
and $\sigma_{EHS}$ by $\bar{r}$, giving 
$a_t=2.33$ and $a_r=0.60$. By employing the analytic expression
for $G_{EHS}(z;\phi_{EHS})$ provided by the Percus-Yevick approximation
\cite{Wertheim:63}, we obtain the values $a_t=3.02$ and $a_r=1.12$,
where the value for $a_r$ in particular is rather
close to the numerical coefficient
in Eq. 
(\ref{hrs.charged}).

Thus, the EHS model suggests that the scaling relations in Eqs. 
(\ref{hts.charged},\ref{hrs.charged}) found from our numerical calculations 
are caused mainly by the leading terms in the series expansions
of the two-body mobility functions. To verify this assertion, we have
included in Figs. \ref{hts.plot} and \ref{hrs.plot} results
for $H^t_s$ and $H^r_s$ obtained by neglecting three-body
contributions and by truncating the two-body series expansions after 
various terms of increasing powers in $(a/r)$. These figures illustrate
our finding that, up to $\phi=0.05$, 
the lowest order contributions to the translational and
rotational two-body mobilities proportional to $r^{-4}$ and
$r^{-6}$, respectively, give by far the most important
contributions to $H^t_s$ and $H^r_s$. 
Higher order two-body terms and the leading-order
three-body term become significant only for $\phi\ge0.05$.
For $H^t_s$, these higher-order terms are of the same signature and  
sum up to increasing deviations in $H^t_s$ from Eq. (\ref{hts.charged}) when
$\phi$ is enlarged beyond $0.05$. With regard to $H^r_s$, however, we observe
a fortuitous partial cancellation between the three-body contribution and
the two-body terms of order ${\cal O}(r^{-8})$, which are
of opposite sign.
As a result, Eq.
(\ref{hrs.charged}) remains valid even up to $\phi\simeq0.15$. 
We mention that this cancellation can also be understood in terms of the
EHS model by reasoning similar to that given above for the leading
two-body contribution to $H^r_s$ \cite{firstone}.

It is further interesting to investigate
how $H^t_s$ and $H^r_s$ are influenced by
added electrolyte. Our corresponding calculations show
a gradual 
transition from the nonlinear scaling relations 
(\ref{hts.charged},\ref{hrs.charged}) to the expressions (\ref{hrs.hs}) and
(\ref{hts.hs.better}) when
the amount of added salt $n_s$ is increased and when the microstructure
changes to a hard-sphere-like structure due to the 
screening of the electrostatic repulsion.

To summarize, we have calculated the translational and rotational short-time
self-diffusion coefficient of charged suspensions by incorporating two-body
and three-body contributions to the HI. As a major result, we have found
for the first time 
substantially different volume fraction dependencies for charged and 
uncharged particles. We were also able to explain the 
observed differences in terms of
an effective hard-sphere model. We mention that recent
depolarized DLS experiments \cite{Bitzer:private} on
deionized suspensions of optically anisotropic particles are in good
agreement with our predicted result for $H^r_s$ in Eq. (\ref{hrs.charged}).
With regard to $H^t_s$, we are not aware of experimental results that are
sufficiently precise at low $\phi$ to distinguish the
$\phi^{4/3}$-behavior from the essentially linear $\phi$-dependence of hard 
spheres. Finally, we point out that interesting qualitative differences between
suspensions of charged particles and hard spheres exist also with respect
to sedimentation \cite{ThiesWeessie:95} and long-time self-diffusion
\cite{Naegele:Baur:forthcoming}.

We are indebted to R. Klein and B. L{\"o}hle 
for useful discussions and to
the Deutsche Forschungsgemeinschaft (SFB 513) for financial support.

\end{document}